\documentclass[journal]{vgtc}                     % final (journal style)

\usepackage{amsfonts}
\usepackage{amsmath}
\usepackage{booktabs}
\usepackage{enumitem}
\usepackage{float}
\usepackage{mathrsfs}
\usepackage{mathtools}
\usepackage{microtype}
\usepackage{multirow}
\usepackage{placeins}

\newcommand{\papertitle}{The Categorical Data Map:\\A Multidimensional Scaling-Based Approach}

\newcommand{\keywordsplain}{Categorical data, dimensionality reduction, cluster analysis, similarity-based representation, information visualization}

\newcommand{\demonstratorlink}{\href{https://dennig.dbvis.de/categorical-data-map}{https://dennig.dbvis.de/categorical-data-map}}

\newcommand{\replicationlink}{OSF (\href{https://osf.io/jzd46}{osf.io/jzd46}) and the Data Repository of the University of Stuttgart (DaRUS) \cite{Dennig2023Replication}}

\newcommand{\orcid}[1]{\hbox{\href{https://orcid.org/#1}{\includegraphics{orcid}}}}

\AtBeginDocument{

}

\newcommand{\subhead}[1]{
  \smallskip
  \noindent\textbf{#1}%
}

\newcommand{\subsubhead}[1]{
  \smallskip
  \noindent\emph{#1}%
}

\newcommand{\compresslist}{
	\setlength{\itemsep}{1pt}
	\setlength{\parskip}{0pt}
	\setlength{\parsep}{0pt}	
}

\DeclareMathAlphabet{\mathpzc}{OT1}{pzc}{m}{it}

\onlineid{1007}

\vgtccategory{Research}

\vgtcpapertype{algorithm/technique}

\title{\papertitle}

\author{%
    \authororcid{Frederik L. Dennig}{0000-0003-1116-8450},
    \authororcid{Lucas Joos}{0000-0001-7049-5203},
    \authororcid{Patrick Paetzold}{0000-0002-1315-4602},
    \authororcid{Daniela Blumberg}{0009-0002-2090-9847},\\
    \authororcid{Oliver Deussen}{0000-0001-5803-2185},
    \authororcid{Daniel A. Keim}{0000-0001-7966-9740}, and
    \authororcid{Maximilian T. Fischer}{0000-0001-8076-1376}
}

\authorfooter{
  \item Frederik L. Dennig, Lucas Joos, Patrick Paetzold, Daniela Blumberg, Oliver Deussen, Daniel A. Keim, and Maximilian T. Fischer are with the University of Konstanz (Germany). E-mail: \{frederik.dennig, lucas.joos, daniela.blumberg, patrick.paetzold, oliver.deussen, keim, max.fischer\}@uni-konstanz.de
}

\abstract{%
  Categorical data does not have an intrinsic definition of distance or order, and thus, established visualization techniques for categorical data only allow for a set-based or frequency-based analysis, e.g., through Euler diagrams or Parallel Sets, and do not support a similarity-based analysis.
We present a dimensionality reduction-based visualization for categorical data based on defining the distance of two data items as the number of varying attributes.
Our technique enables users to pre-attentively detect groups of similar data items and observe the properties of the projection, such as attributes strongly influencing the embedding.
Our prototype visually encodes data properties in an enhanced scatterplot-like visualization, visualizing attributes in the background to show the distribution of categories.
We propose two graph-based measures to quantify the plot's visual quality for ranking attributes according to their contribution to cluster cohesion.
To demonstrate the capabilities of our method, we compare it to Euler diagrams and Parallel Sets regarding visual scalability and evaluate it quantitatively on seven real-world datasets using a range of common quality metrics.
We conducted an expert study with five data scientists analyzing the Titanic and Mushroom datasets with up to 23 attributes and 8124 category combinations.
Our results indicate that our Categorical Data Map is an effective analysis method for large datasets with a high number of category combinations.
}


\teaser{
\centering
\includegraphics[width=\linewidth]{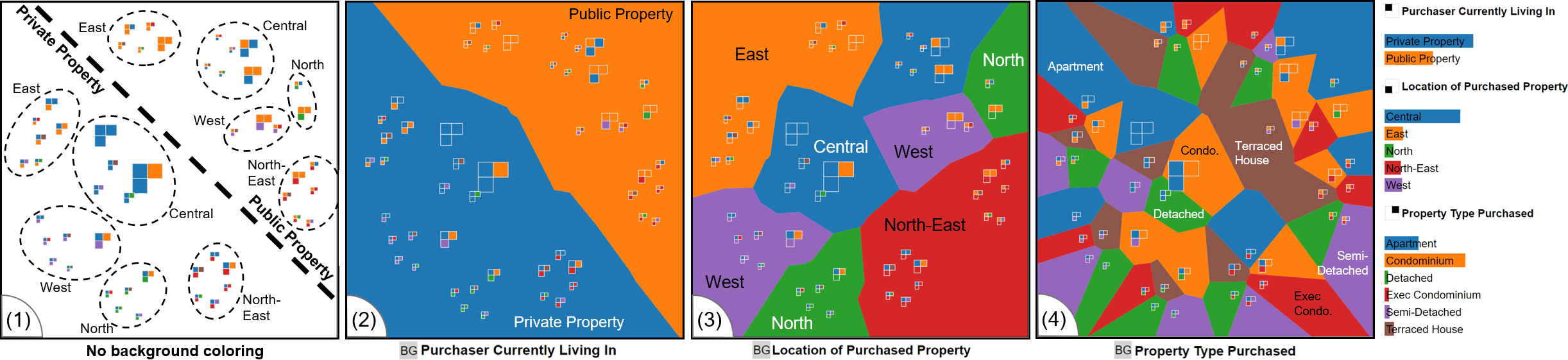}
\caption{
The \emph{Categorical Data Map} enables projection-based analysis of categorical data here exemplified by the \emph{Property Sales} dataset~\cite{Koh2011} with \emph{MDS} \cite{Maaten2008} using the \emph{Jaccard coefficient} \cite{Jaccard1912}: (1) shows 10 groups without layout enrichment. 
Our method reveals the patterns annotated in (1) in plots (2)-(4). 
(2) shows a clear separation between \textrm{Private Property} vs \textrm{Public Property}.
(3) indicates boundaries and symmetries for the \textrm{Location of Purchased Property} attribute, while in (4), the \textrm{Property Type Purchased} contributes the least to the clusters.
The glyph sizes encode the subset sizes, revealing that categories \textrm{Private Propriety} and \textrm{Central} often occur together.
}
\label{fig:basic-example}
}

\graphicspath{{figs/}{figures/}{pictures/}{images/}{./}} % where to search for the images

\usepackage{tabu}                      % only used for the table example
\usepackage{booktabs}                  % only used for the table example
\usepackage{lipsum}                    % used to generate placeholder text
\usepackage{mwe}                       % used to generate placeholder figures

\usepackage{mathptmx}                  % use matching math font

\begin{document}

%%%%%%%%%%%%%%%%%%%%%%%%%%%%%%%%%%%%%%%%%%%%%%%%%%%%%%%%%%%%%%%%
%%%%%%%%%%%%%%%%%%%%%% START OF THE PAPER %%%%%%%%%%%%%%%%%%%%%%
%%%%%%%%%%%%%%%%%%%%%%%%%%%%%%%%%%%%%%%%%%%%%%%%%%%%%%%%%%%%%%%%

\firstsection{Introduction}

\maketitle

%\section{Introduction}
%\label{sec:introduction}

Categorical data can be encountered in numerous domains, such as representing inventory data describing product properties like color in sales or bioinformatics, encoding the genes formed by nucleotide sequences \cite{Agresti2018}.
In contrast to numeric and ordinal data, categorical data does \emph{not} have an intrinsic order or distance associated with each value pair.
The visual analysis of categorical data is challenging since categorical data describes an attribute by name only, with the only supported operators being \emph{equality}, \emph{set membership}, and \emph{mode}.
Currently, there are two widespread methods of visualizing categorical data:
(1) \emph{Frequency-based visualizations}~\cite{Hofmann2000,Wittenburg2001,Spenke2003} map the categorical values to their frequencies, for example, through bar charts, pie charts, or enhanced variants, such as stacked bar charts.
In contrast, (2) \emph{set visualizations} solely focus on the set nature of categorical data items, specifically their intersections~\cite{Alsallakh2016}.
Examples include such as Euler diagrams \cite{Paetzold2023} and UpSet plots \cite{Lex2014}.
Set visualizations like Euler diagrams do not scale well for sets with many intersections because visual clutter is detrimental to their readability.
Other, less common solutions treat dimensions independently and map data to 
a continuous design model~\cite{Teoh2003,Jerding1998,Rosario2004}, leveraging visualization types that initially have been designed for numerical data, such as scatterplots or parallel coordinate plots.
However, these approaches deviate from the \emph{discrete} nature of categorical data and suffer from visual clutter and overplotting, limiting their readability~\cite{Kosara2006}.
Approaches, such as Parallel Sets~\cite{Bendix2005} and Sankey diagrams~\cite{Sankey1898}, follow the frequency and set-based paradigms.
These approaches trade effectiveness in visualizing the presence of small subsets for the presentation of frequency information.
These approaches require additional design considerations since they tend to emphasize preselected attributes over others~\cite{Dennig2021}.

None of the previously described techniques support the similarity-based analysis of categorical data, i.e., deriving the \emph{similarity} of categorical data items as distances such that similar data items are placed close to each other while differing data items are positioned far apart.
Analyzing categorical data based on a group or subset similarity is useful, e.g., visually clustering data items only differing in a few attributes can help us better understand important characteristics of the group.
Generally, this would allow us to apply methods from cluster analysis to categorical data.
We follow the suggestion by Broeksema et al. \cite{Broeksema2013} to investigate multidimensional scaling to generate \emph{visual mappings} that enable the interpretation of distances and simultaneously convey the properties of data items, i.e., effectively visualizing an item's attributes by using color and position to visually encode attributes.
Through this, we address the combinatorial problem of categorical data, i.e., that with the increasing number of attributes and categories, the number of required colors to represent a category with distinguishable colors becomes increasingly difficult.
\noindent
Tackling these challenges, we contribute the following:
\begin{enumerate}[label=(\arabic*),left=0pt]
\compresslist
\item A technique applying multidimensional scaling to categorical data while \emph{visually encoding} the category distribution into the background. Through \emph{layout enrichment}, we enable the exploration of the category distribution, enhancing orientation and navigation.
Additionally, we contribute \emph{four glyph designs} to represent categorical subsets.
\item \emph{Quality measures} based on subset distribution to guide the analysis, recommending layout enriched views on attributes contributing strongly to clusters and subset separation.
\item A \emph{quantitative comparison} to multiple correspondence analysis-based projections and a \textit{qualitative expert study} validating the effectiveness of our approach. 
\item An \emph{online demonstrator} (\demonstratorlink) making the acquired results accessible.
To further aid reproducibility, we \emph{openly publish} all our \emph{datasets} and \emph{source code} via  \replicationlink.
\end{enumerate}

With this work, we aim to widen the analytical capabilities for categorical data, particularly for exploratory analysis.

\section{Related Work}
\label{sec:related-work}

Our approach is related to visualization and dimensionality reduction methods for categorical data.
Furthermore, we propose a layout enrichment for multidimensional projections and contribute visual quality metrics for categorical data projections.

\subsection{Visualization Techniques for Categorical Data}
\label{sec:related-work:visualization-techniques}

\emph{Set visualization} is one of the core techniques for categorical data.
To visualize the members of sets and their intersections, Venn and Euler diagrams are the two most prevalent representations~\cite{Baron1969}.
Multiple adaptations of both techniques mitigate challenges, e.g., to preserve semantics~\cite{Kehlbeck2022}, draw area-proportional diagrams~\cite{Perez2018},
or incorporate glyphs to show additional information~\cite{Micallef2012}.
Other set visualization techniques use lines to indicate set intersections~\cite{Rodgers2015} and matrices to show the cardinality of intersection sets~\cite{Lex2014Upset},
or include the semantic context to visualize sets~\cite{Meulemans2013}.
Alsallakh et al. presented a comprehensive survey on set visualizations~\cite{Alsallakh2016}.
There are also \emph{frequency-based visualization} methods that focus on attribute frequencies, such as Mosaic plots~\cite{Hofmann2000} and Parallel Bargrams~\cite{Wittenburg2001} by mapping data item occurrences to one or multiple attributes, e.g., a rectangle's area.
Other methods map data to a continuous design model, such that they are compatible with visualization for numeric data, e.g., Rosario et al.~\cite{Rosario2004} describe the mapping of categorical data to numeric values for the visualization in Parallel Coordinates~\cite{Inselberg85}.
Hybrid methods consider both aspects, e.g., Parallel Sets~\cite{Bendix2005,Kosara2006} and Sankey diagrams~\cite{Sankey1898}.
However, Parallel Sets and Sankey diagrams can suffer from the Müller-Lyer and Sine illusions~\cite{Day1991,VanderPlas2015} where lines seem to vary in distance or length, affecting the accurate interpretation of frequencies and proportions.

While plenty of approaches visualize categorical data, to the best of our knowledge, none allows identifying groups of similar data items.
Thus, we propose a visualization that focuses on similarity.

\subsection{Dimensionality Reduction for Categorical Data}
\label{sec:related-work:dimensinality-reduction}

Our approach makes use of dimensionality reduction (DR). 
However, there exist DR methods for categorical data that do not focus on similarity but rather describe the central oppositions in the data~\cite{Greenacre2006}.
When needing to reduce the dimensionality of categorical data, Correspondence Analysis (CA), similar to Principal Component Analysis (PCA)~\cite{Jolliffe1986} for numerical data, extracts the standard coordinates, yielding a Biplot~\cite{Gabriel1971} of the reduced space.
In case of more than two categorical variables, Multiple Correspondence Analysis (MCA) can be used to reduce the number of dimensions showing the central oppositions~\cite{Greenacre2006}. 
Factor analysis of mixed data (FAMD) is a principal component technique for continuous and categorical variables~\cite{Pages2014}.
The continuous variables are scaled to unit variance, and the categorical variables are transformed into a disjunctive data table and then scaled using the specific scaling of MCA to balance the influence of both continuous and categorical variables in the analysis. 
Multiple Factor Analysis (MFA) combines these methods for mixed data:
It uses PCA when variables are quantitative, MCA when variables are qualitative, and FAMD when the active variables belong to both of the two types.
The Data Context Map \cite{Cheng2016} visualizes \emph{mixed-data} using an MDS-based plot and displays categorical attributes on top of the projection while also coloring points and regions according to the predominant category.
The approach by Thane et al. \cite{Thane2023} uses force-directed graph layouts to visualize categorical datasets representing categories as nodes while edges represent their co-occurrence.
MCA can embed categorical data but, like PCA, is a linear dimensionality reduction technique and thus not able to detect non-linear relationships \cite{Greenacre2006,Broeksema2013}.
We propose using MDS to visualize the similarity of categorical data points in a scatterplot-like layout.

\subsection{Layout Enrichment for 2-Dimensional Data Projections}
\label{sec:related-work:layout-enrichment}

The idea to enrich scatterplot layouts by encoding additional information in the background of a projection is not new~\cite{Nonato2019}.
The main usage occurs for the visualization of distortions in the topology of the embedding resulting from DR \cite{Aupetit2007}.
The following approaches make use of Voronoi diagrams \cite{Aurenhammer1991} to encode additional information in the background of a projection.
Lespinats and Aupetit proposed CheckViz~\cite{Lespinats2011}, visualizing the presence of \emph{tears} (i.e., missing neighborhood) and \emph{shuffled data} (i.e., wrong neighborhood).
Broeksema et al. explored the visualization of categorical data, combining MCA with an enhanced treeview to integrate data record information visualizing user-selected categories.
However, they did not address the high redundancy of categorical datasets~\cite{Broeksema2013}.
Sohns et al. followed a similar approach; however, they used non-linear DR methods to project \emph{mixed data} while using categorical attributes to highlight areas of the embedding space.
However, this approach excludes all categorical attributes from the DR process altogether~\cite{Sohns2022}.
DICON enables the analysis of multidimensional clusters with an interactive icon-based visualization that encodes additional statistical information visually using space-filling methods, including Voronoi diagrams \cite{Cao2011}.
Aside from using Voronoi diagrams, other methods for layout enrichment exist \cite{Blumberg2024}.
Morariu et al. encode the projection's quality into the plot's background using contours showing the embedding of projections called the metamap \cite{Morariu2021}.
Layout enrichment methods largely focus on visualizing distortions of the projection. The approach by Broeksema et al. \cite{Broeksema2013} does not address the analysis of a single attribute, so we propose a new enrichment that encodes the category of an attribute using color.

\subsection{Metrics for Quality and Patterns in Visualizations}
\label{sec:related-work:measures}

Quality metrics for visualizations describe a set of measurements designed to optimize visualizations in terms of \emph{readability} and \emph{clutter reduction}~\cite{Behrisch2018}.
Other metrics quantify the presence of \emph{patterns} in a visualization.
Instead of measuring quality, pattern metrics can be used to compare and rank different visualizations based on specific properties.
Examples are: Magnostics for matrix visualizations~\cite{Behrisch2017}, Scagnostics for general patterns and trends on scatterplots of numeric data~\cite{Wilkinson2005},
Pargnostics for parallel coordinate plots~\cite{Dasgupta2010}, Visualgnostics for projections of high-dimensional data~\cite{Lehman2015}, Pixgnostics for pixel-based visualizations~\cite{Schneidewind2006}, and ParSetgnostics for Parallel Sets~\cite{Dennig2021}.
SepMe is a machine-learning-based approach to quantify the presence of clusters in scatterplots~\cite{Aupetit2016}, while ClustMe quantifies the visual separation of classes in scatterplots~\cite{Abbas2019}.
Aupetit and Catz \cite{Aupetit2005} addressed the analysis of high-dimensional labeled data using graphs, including Voronoi diagrams. However, this approach does not address categorical data analysis, i.e., where no numerical attributes are present. 

We contribute two novel measures for quantifying visual quality for 2-dimensional projections of categorical data.
In this way, we improve the exploration of categorical data by recommending layout-enriched views according to their visual structure.

\section{Constructing the Categorical Data Map}
\label{sec:method}

Typically, categorical datasets exhibit inherent \emph{sparsity}, i.e., only a fraction of all possible category combinations is present in a dataset, e.g., for the Mushroom dataset, only 8124 out of 243.799.621.632.000 possible combinations.
Thus, we assume that there are relationships among the existing categories restricting their combinations.
Additionally, categorical datasets can be \emph{highly redundant}, e.g., the Titanic dataset contains 2201 data items but only 24 unique entries, i.e., all data items can be assigned to one of 24 subsets.
Thus, we focus on categorical subsets as subsets of unique attribute values.
These subsets are our main representations, enabling us to assign a frequency.
We leverage these properties in the design of the \emph{Categorical Data Map} as an analytical approach for the similarity-based analysis of categorical subsets with the following \emph{constraints}:
\begin{enumerate}[label=(C\arabic*),left=0pt]
\compresslist
\item Distances of categorical subsets in a scatterplot should indicate similarity, i.e., subsets with a smaller distance should differ in fewer attributes than subsets with a larger distance.
\item Allow analysts to find groups of subsets by clustering similar categorical subsets and separating outliers.
\item Highlight attributes contributing to the clustering of subsets enabling navigation and orientation in the projection.
\item Provide a recommendation for attributes to explore first, linked to the distribution of categories in the plot.
\end{enumerate}

An example of our approach is shown in \autoref{fig:basic-example}.
(C1) and (C2) are described further in \autoref{sec:method:projecting}.
We address (C3) by evaluating different glyph designs and layout enrichments for subsets of categorical data (see \autoref{sec:method:glyphs}). 
We address (C4) in \autoref{sec:method:measure}, describing measures to rank attributes according to their degree of splitting the embedding into connected areas.
In the following, we describe how we derive distance relations of categorical data and how a projection-based approach, i.e., the \emph{Categorical Data Map}, is constructed. 

\subsection{Projecting Categorical Data}
\label{sec:method:projecting}

The \textit{Categorical Data Map} enables the visual clustering of similar categorical subsets and separating outliers, addressing (C1) and (C2).
At the core, we rely on DR to create a scatterplot-like visualization.
In general, we describe encoding $E$, distance measure $M$, DR method $P$, and overlap reduction method $O$ to project a categorical dataset $x$ by applying $O(P(M(E(x))))$.

\subhead{Encoding (E):}
We convert all data items into a set representing their categorical data values.
We define the set of all attributes as
$\mathcal{A} \coloneqq \{a_1, a_2, \dots, a_{\left|\mathcal{A}\right|}\}$ %
and the possible categories associated with attribute $a_i$ as the set %
$\mathcal{C}_{i} \coloneqq \{c_i^1,c_i^2, \dots, c_i^{\left|\mathcal{C}_i\right|} \}$ with $i \in \mathbb{N}$.
$\left|\mathcal{A}\right|$ is the cardinality of a set representing a data item, i.e., the number of attributes since a data item has one category associated with each attribute.
We denote a data item as $x_n = (c_1^{n_1} , c_2^{n_2} ,...,c_{\left|\mathcal{A}\right|}^{n_{\left|\mathcal{A}\right|}})$. 
From a practical point of view, we make sure that all categories have a unique descriptor across all attributes.
We then create a representation compatible with the distance measure.
We explored the \textit{set representation} and two variants of \textit{one-hot encoding} \cite{Brownlee2017, Hancock2020} (see supplementary material for more details).
\subhead{Distance Measure (M):} With the set representation, we can describe the categories of a data item to define similarity.
Based on surveys on distance measures for categorical data \cite{Cha2007, Boriah2008, Sulc2019}, we chose and evaluated three set-based distance measures: \textit{Overlap coefficient} \cite{Waggener1994}, \textit{Jaccard Similarity Index} \cite{Jaccard1912}, and \textit{Sørenson-Dice coefficient} \cite{Sorenson1948}.
By including one-hot encoding, converting each categorical value to a new binary dimension enables us to use classical distance measures, such as \textit{Euclidean} or \textit{Manhattan distance}, to describe a dissimilarity relationship (see supplementary material for more details).

\subhead{Projection Method (P):}
DR techniques are a set of non-/linear transformation methods with which a dataset's dimensionality can be reduced.
We compared the following two DR methods.

\subsubhead{Multiple Correspondence Analysis (MCA):}
This method is the categorical equivalent of PCA.
MCA creates groups of items that are similar according to their categories.
Objects sharing the same categories are placed close together, and objects with differing categories are placed far apart~\cite{Greenacre2006}.
To our knowledge, MCA is the only existing technique that directly uses the set representation of categorical data.
\subsubhead{Multidimensional Scaling (MDS):}
This method describes a set of linear and nonlinear DR techniques that attempt to preserve pairwise distances.
Multiple criteria are possible; Kruskal’s stress optimization criterion is usually used~\cite{Kruskal1964}.
We create a dissimilarity matrix to compute the projection given one of the described distance measures.

\smallskip

We chose the two methods based on their popularity and common usage in visual data analysis~\cite{Broeksema2013,Dennig2023Fsds} and compared MDS and MCA as DR methods for categorical data.

However, a key difference between both methods is that MCA reduces the number of projected points to the number of unique subsets, while MDS, applied naively, would result in a number of projected points equal to the number of categorial data items.
Since categorical datasets can contain many duplicates, projecting each data point individually and using a DR method for numeric data (e.g., MDS) could lead to multiple data points being projected to the same position.
The main reason is that the distance of identical points is zero.
To achieve a comparable result, i.e., the same number of projected points, we remove all duplicates and project one data point for each unique combination of attribute values, i.e., for each categorical subset, describing the prototype of the represented data subset.
A second reason is that we want to show the subsets represented by a point irrespective of the method (e.g., MCA or MDS).
We visually represent a subset's size (see~\autoref{sec:method:glyphs}).
Reducing the number of data points also improves the runtime of projection algorithms for datasets with duplicate items.

\subhead{Overlap Reduction (O):}
Given that some subsets in the categorical data may differ in only one or a few attributes, these subsets will be projected close to each other.
This property is desirable in the design of a map by keeping the distances representing similarity coherent.
However, it may also introduce overlap if the projected point visually encodes the subset categories through a glyph representation.
Additionally, points that are close together will yield small or narrow-shaped Voronoi cells. 
Thus, we allow users to reduce the overlap after projecting the data using a method based on force-directed graph drawing.
This type of layout applies forces to the nodes and edges of a graph~\cite{Kobourov2012}.
We add a repulsive force to all points with a strength equal to the radius of the glyph while all points are vertices of a fully connected graph, forcing all points into a configuration without overlap but with minimal space in between the glyphs.

\subsection{Representing Categorical Data Subsets in Scatterplots}
\label{sec:method:glyphs}

We implemented the visual components of the \emph{Categorical Data Map} using D3~\cite{Bostock2011}.
To represent categorical subsets, we developed four glyph representations and the layout enrichment based on experiences gained during the design phase, addressing (C3).
To represent categories, we use the \texttt{d3.schemeCategory10} color scale, a well-established color scale for categorical data.
\begin{figure}%[tbh]
%\centering
\raisebox{0.03\height}{\includegraphics[width=2.0cm]{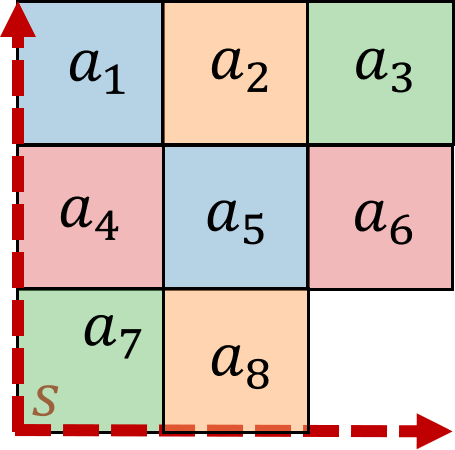}}
\hspace{0.05cm} 
\includegraphics[width=2.0cm]{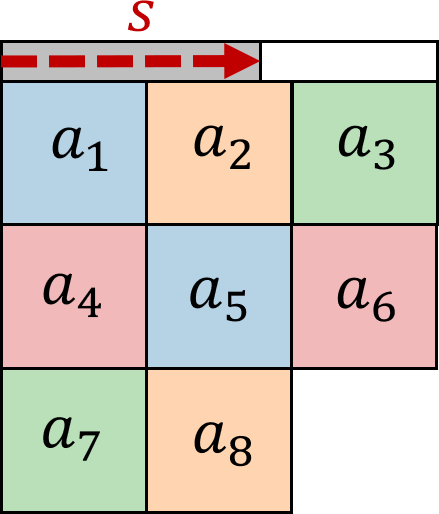}
\hspace{0.05cm} 
\raisebox{0.03\height}{\includegraphics[width=2.0cm]{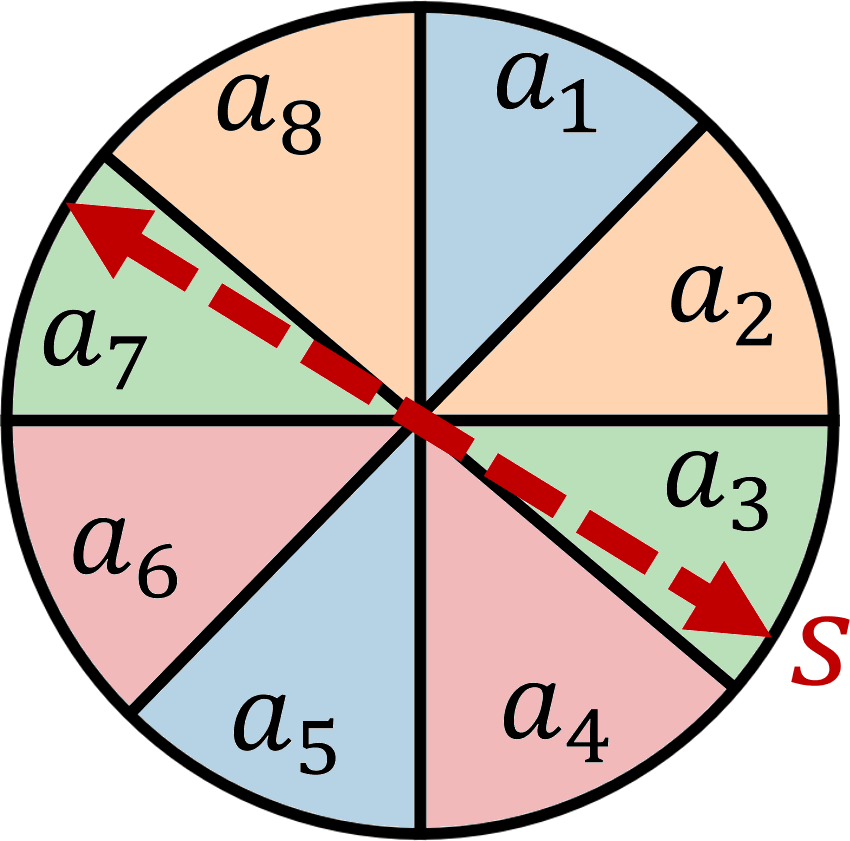}}
\hspace{0.05cm} 
\raisebox{0.03\height}{\includegraphics[width=2.0cm]{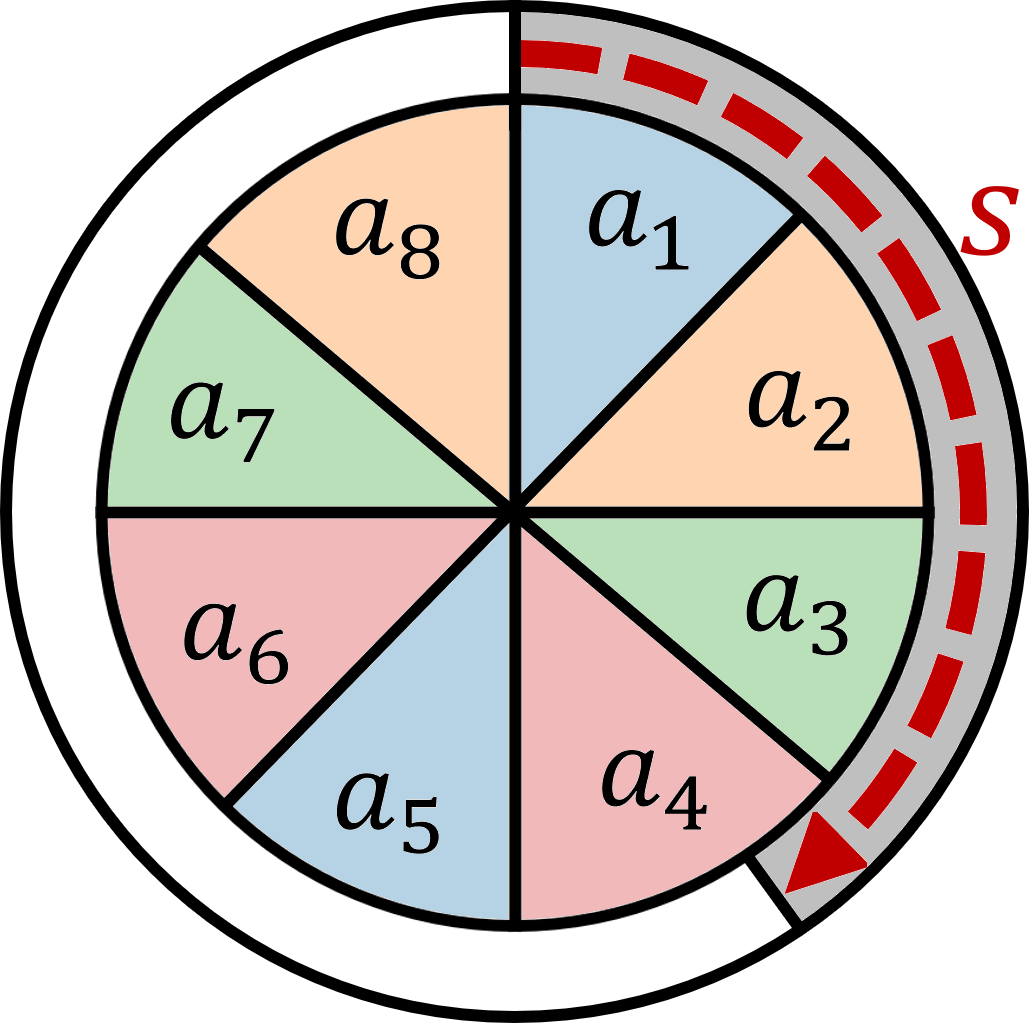}}\\
\footnotesize{
\hspace*{0.15cm}(a) Area square\hspace{0.45cm}(b) Bar square\hspace{0.6cm}(c) Area circle\hspace{0.7cm}(d) Arc circle
}
\caption{
Representation of subsets for a dataset with eight attributes.
(a)~shows the eight attributes in four segments with the same area while the size encodes the overall subset size.
(b)~shows a similar glyph, but instead, the size is encoded by a bar at the top, and all glyphs have the same size.
(c)~Encodes the attributes similar to the area square but is circle-shaped.
(d)~encodes the size by an arc filled according to the subset size.
}
\label{fig:glyphs}
\end{figure}
\subhead{Glyph Representation:}
To represent categorical subsets, we developed four glyph representations.
All glyphs visualize the attributes and their respective values by dividing a square or circle into segments of equal size, such that each segment represents one attribute.
This square-based glyph is inspired by pixel visualizations pioneered by Keim et al.~\cite{Keim2000}.
In~\autoref{fig:glyphs}, this is represented by the categories $a_1$ to $a_8$ for the case of a dataset with eight attributes.
For all glyphs, the segments are colored according to the respective category of the attribute. 
However, we discuss some limitations in~\autoref{sec:discussion}.
The area-based glyphs represent the relative size of a subset $s \in \mathbb{N}$ by the area (see \autoref{fig:glyphs} (a) and (c)).
Thus, we calculate the width and height accordingly.
The bar- and arc-based glyphs have a fixed size to minimize space requirements and overlap issues with neighboring glyphs (see \autoref{fig:glyphs} (b) and (d)).
To reduce overlap while preserving the relative proximity of the projected points, we decided to map a subset's size $s \in \mathbb{N}$ to a bar at the top or an arc surrounding the glyph as an alternative encoding for the subset size.
Hence, each unique subset is represented by a square or circle sized relative to the percentage of data points the subset represents or an indicator filled accordingly.
This enables users to perceive similar subsets and assess the size of each group.

\subhead{Layout Enrichment:}\label{sec:topology}
To enable the observation of cluster characteristics and explore attributes in the projected space, we show a Voronoi diagram \cite{Aurenhammer1991} for a selected attribute (see \autoref{fig:basic-example}).
The Voronoi diagram automatically partitions the map into polygons such that each polygon contains exactly one subset.
By selecting one attribute of interest, the partition for the selected attribute gets displayed in the background of the projection.
The color of the polygon then encodes the category of the selected attribute.
Thereby, it is possible to directly spot cluster regions for the selected attribute and to identify cluster boundaries and outlying data points.
The appearance of the background can differ a lot across attributes (see ~\autoref{fig:fracturedness}).
Attributes form distinct contiguous areas of different sizes, indicating a neighborhood or larger area of subsets of the same category.
We added \emph{detail-on-demand} using tooltips, allowing users to see the respective category for each polygon directly.

\subsection{Measuring Fracturedness}
\label{sec:method:measure}

\begin{figure}[t]
\includegraphics[width=\linewidth]{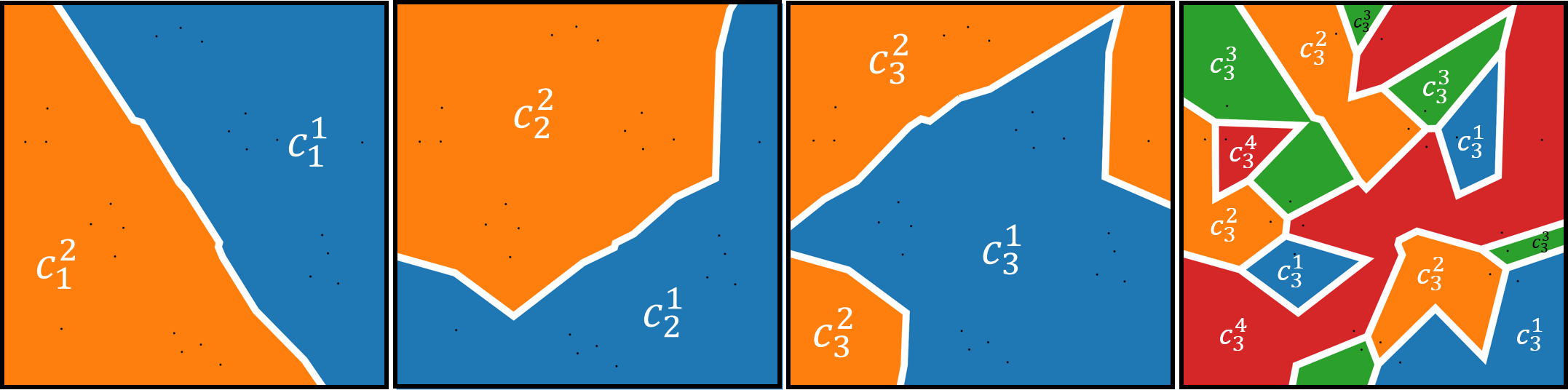} \\
{\footnotesize
$\mathcal{F}_\textrm{edge}(a_1) = 0.17$\hspace{0.07cm}
$\mathcal{F}_\textrm{edge}(a_2) = 0.17$\hspace{0.04cm}
$\mathcal{F}_\textrm{edge}(a_3) = 0.42$\hspace{0.04cm} $\mathcal{F}_\textrm{edge}(a_4) = 0.79$\hfill\\
$\mathcal{F}_\textrm{comp}(a_1) = 0$\hspace{0.36cm} $\mathcal{F}_\textrm{comp}(a_2) = 0$\hspace{0.36cm} $\mathcal{F}_\textrm{comp}(a_3) = 0$\hspace{0.36cm} $\mathcal{F}_\textrm{comp}(a_4) = 0.69$\hfill
}
\caption{The \emph{fracturedness} of attributes differs a lot and can imply an order, i.e., increasing from left to right. The examples are derived from the Titanic dataset~\cite{Titanic95}. The edge-based (i.e., $\mathcal{F}_\textrm{edge}$) and component-based fracturedness (i.e., $\mathcal{F}_\textrm{comp}$) values are provided below for each attribute.}
\label{fig:fracturedness}
\end{figure}
We quantify \emph{fracturedness}, generally defined as the strength with which the Voronoi partitioning of an attribute appears disjointed and fractured (see \autoref{fig:fracturedness}).
We use \emph{fracturedness} to suggest attributes for analysis, e.g., the lower the fracturedness value, the larger the contiguous areas of categories and thus the more straightforward to orient along, addressing (C4).
We use the Delaunay triangulation of the Voronoi diagram \cite{Aurenhammer1991} as a basis for our measures.
In contrast to Aupetit and Catz \cite{Aupetit2005}, we describe measures for purely categorical datasets.
Before describing the measures, we define the common notations following established notations \cite{Clark1991, Aupetit2005}.
Let $G \coloneqq (V, E)$ be the Delaunay triangulation of the discrete set of points $P$ resulting from the projection (see~\autoref{sec:method:projecting}).
Thus, $G$ is an undirected graph and the dual graph of the Voronoi diagram of the points $P$.
Therefore, there exists exactly one $v \in V$ for every $p \in P$ defining its x,y-location and categories. 
Each vertex $v \in V$ has exactly one associated category $\mathscr{C}_n(v) \in \mathcal{C}_n$ for each attribute $a_n \in \mathcal{A}$. 

\begin{figure}[t]
\centering
\includegraphics[width=\linewidth,trim=0 75 0 65,clip]{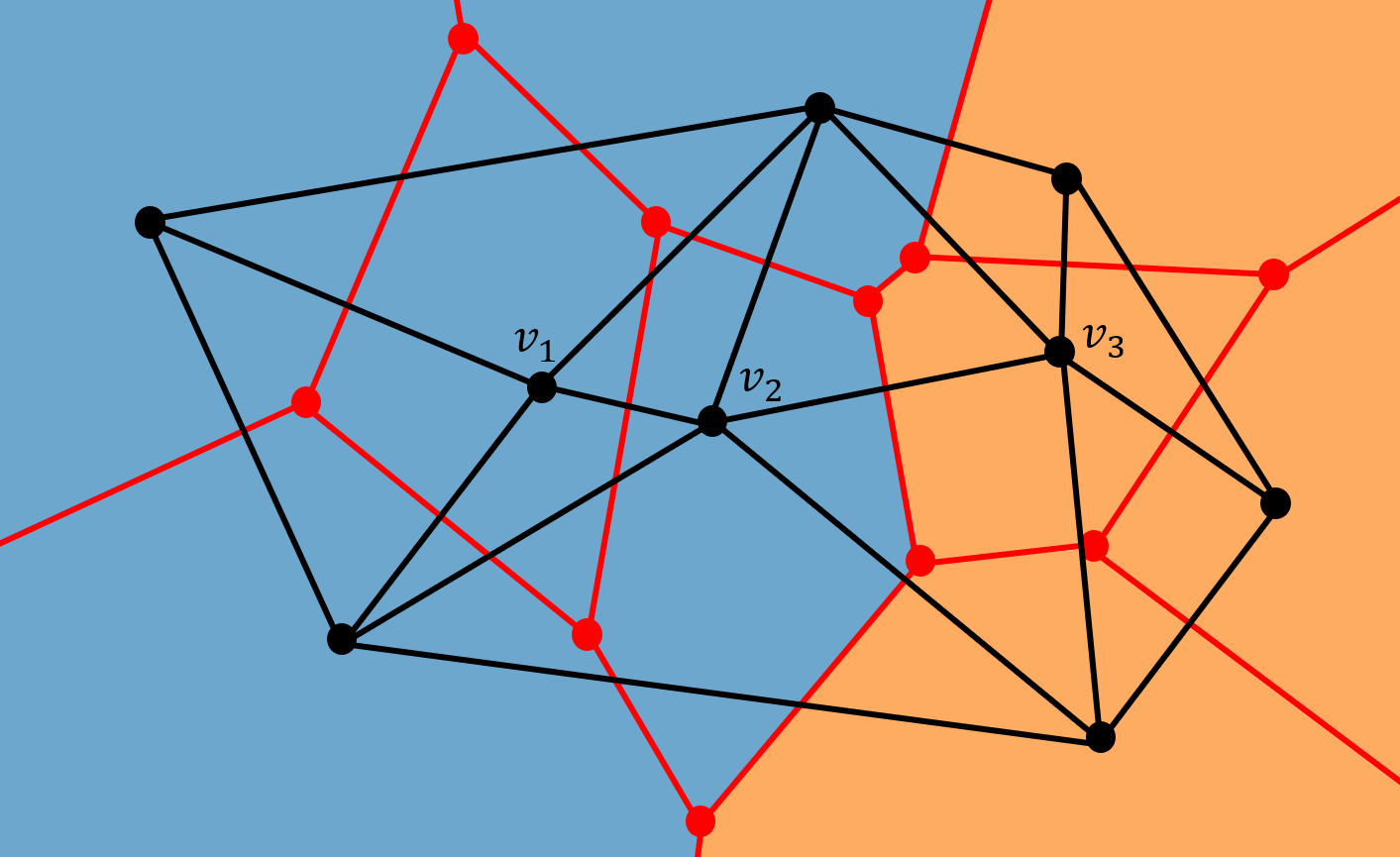}
\caption{
We illustrate \emph{edge-based fracturedness} with a Delaunay triangulation shown in black, and a Voronoi partitioning with cell borders shown in red.
The cells are colored according to the categories of an attribute.
$v_1$, $v_2$ and $v_3$ are vertices of the Delaunay triangulation.
The edge ${v_1, v_2}$ will not contribute to edge-based fracturedness, since it connects cells representing the same category of a given attribute.
Edge ${v_2, v_3}$ contributes to edge-based fracturedness because it connects cells representing different categories.
}
\label{fig:delaunay-1}
\vspace*{-1em}
\end{figure}

\subhead{Edge-based Fracturedness:}
We measure the number of edges in $G$ that connect cells with different associated attributes.
This concept is shown in~\autoref{fig:delaunay-1}.
We define an edge $e \in E$ as $\{v_1, v_2\}$ with $v_1, v_2 \in V$ and $v_1 \neq v_2$.
An edge contributes to \emph{fracturedness}, if the category for the analyzed attribute $a_n$ and its associated categories in $\mathcal{C}_n$ differ for the connected vertices, i.e., $\mathscr{C}_n(v_1) \neq \mathscr{C}_n(v_2)$ for $\{v_1, v_2\} \in E$.
\textit{Edge-based fracturedness} is defined as $\mathpzc{F}_{edge} : \mathcal{A} \mapsto [0,1]$ and calculated using \autoref{eq:fracturedness-edge}.
\begin{equation}\label{eq:fracturedness-edge}
\mathpzc{F}_{edge}(a_n) \coloneqq \frac{\left| \{ v_1, v_2 \} \in E : \mathscr{C}_n(v_1) \neq \mathscr{C}_n(v_2) \right|}{|E|} \text{ with } a_n \in \mathcal{A}
\end{equation}

\begin{figure}[t]
\centering
\includegraphics[width=\linewidth, trim=20 70 10 45,clip]{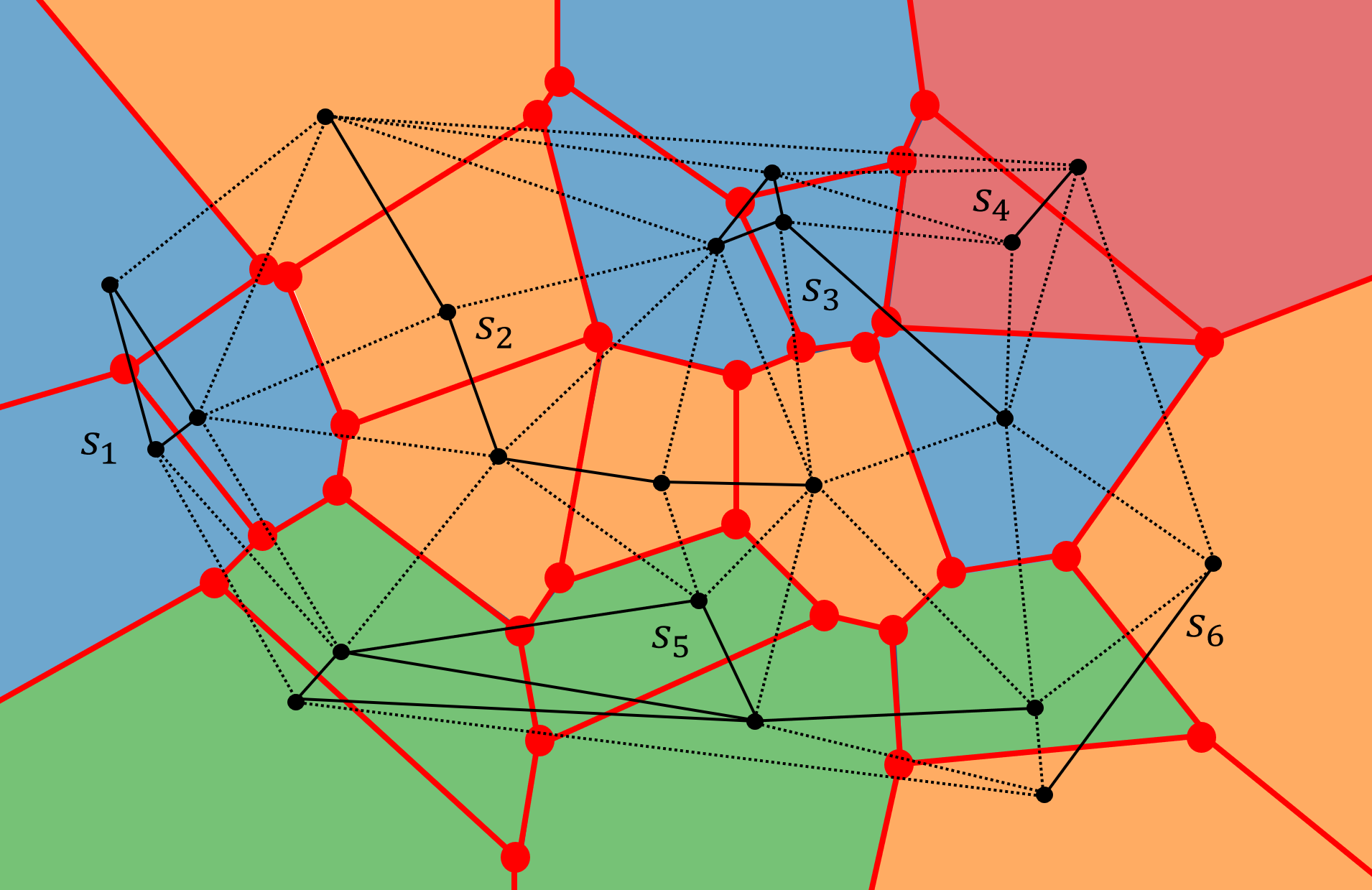}
\caption{
We describe \emph{component-based fracturedness} with a Voronoi partitioning with cell borders shown in red.
The associated Delaunay triangulation is shown in black.
The cells are colored according to the categories of an attribute.
$s_1$ to $s_6$ are six components induced by an attribute through the subgraphs associated with a category.
Solid lines connect each subgraph, while dashed lines are not part of any subgraph.
With six components $\mathpzc{F}_{comp} = 0.33$ for the attribute (see \autoref{eq:fracturedness}).
}
\label{fig:delaunay-2}
\end{figure}

\subhead{Component-based Fracturedness:}
This measure quantifies the number of continuous areas an attribute produces in the plot through its categories.
We show the concept of \emph{component-based fracturedness} in~\autoref{fig:delaunay-2}.
Each category $c \in \mathcal{C}_{n}$ defines an induced subgraph $G[S(c)]$ of $G$, with $S(c) \subset V$ for all $c \in \mathcal{C}_n$ of an attribute $a_n \in \mathcal{A}$.
The induced subgraph $G[S(c)]$ is a graph with the vertices $S(c)$ and the edges in $E$ with both of its vertices in $S(c)$.
We formally define $S(c)$ for a category $c \in \mathcal{C}_n$ in~\autoref{eq:vertex-subset}.
\begin{equation}\label{eq:vertex-subset}
S(c) \coloneqq \{v \, | \, v \in V, \mathscr{C}_n(v) = c\} \text{ for } c \in \mathcal{C}_n \text{ of } a_n \in \mathcal{A}
\end{equation}
With this definition, a category defines a partition of $V$, i.e., $\bigcup_{c \in \mathcal{C}_n} S(c) = V$ and a vertex $v \in V$ can only have one category $\mathscr{C}_n(v)$, thus $\bigcap_{c \in \mathcal{C}_n} S(c) = \emptyset$ for a given attribute $a_n$.
Therefore, there exits $\left|\mathcal{C}_n\right|$ subgraphs of $G$ for attribute $a_n \in \mathcal{A}$.
Let $\omega(G)$ be the number of connected components of any graph $G$.
The \emph{component-based fracturedness} is dependent on the number of connected components of all subgraphs $\omega(G[S(c)])$ for each $c \in \mathcal{C}_n$ (see $s_1$ to $s_6$ in \autoref{fig:delaunay-2}).
We define the sum of the number of components of all induced subgraphs as $\Omega(a_n)$ for an attribute $a_n \in \mathcal{A}$.
$\Omega(a_n)$ is formally defined in~\autoref{eq:all-components}:
\begin{equation}\label{eq:all-components}
\Omega(a_n) \coloneqq \sum\limits_{c \in \mathcal{C}_n}\omega(G[S(c)]) \text{ with } a_n \in \mathcal{A}
\end{equation}
We can also quantify the fracturedness a single category contributes to the overall measure.
This allows us to differentiate categories forming contiguous areas and highly fractured ones.
The fracturedness $\mathpzc{f}_{\textrm{comp}}(c)$ of a single category $c \in \mathcal{C}_n$ is defined in \autoref{eq:fracturedness-single}:
\begin{equation}\label{eq:fracturedness-single}
\mathpzc{f}_{\textrm{comp}}(c) \coloneqq \frac{\omega(G[S(c)]) - 1}{\Omega(a_n)} \text{ with } c \in \mathcal{C}_n \text{ of } a_n \in \mathcal{A} 
\end{equation}
\textit{Component-based fracturedness} is defined as $\mathpzc{F}_{comp} : \mathcal{A} \mapsto [0,1]$ and calculated using \autoref{eq:fracturedness}.
It allows us to compare different attributes and is an alternative measure to $\mathpzc{F}_{edge}(a_n)$.
\begin{equation}\label{eq:fracturedness}
\mathpzc{F}_{\textrm{comp}}(a_n) \coloneqq  1 - \frac{\left|\mathcal{C}_n\right|}{\Omega(a_n)} \text{ with } a_n \in \mathcal{A}
\end{equation}
The sum of all component-based fracturedness values of individual categories $c \in \mathcal{C}_n$ is equal to the fracturedness of the attribute $a_n \in \mathcal{A}$.
We express this relationship in \autoref{eq:fracturedness-sum}:
\begin{equation}\label{eq:fracturedness-sum}
\mathpzc{F}_{\textrm{comp}}(a_n) = \sum\limits_{c \in \mathcal{C}_n}\mathpzc{f}_{\textrm{comp}}(c) \text{ with } a_n \in \mathcal{A}
\end{equation}
A mathematical proof of the equivalence described in \autoref{eq:fracturedness-sum} can be found in the supplementary material.

\subsection{Interacting with Attributes and Subsets}

Our prototype allows interactions on the attributes of the dataset shown in the side panel and projected subsets.

\begin{figure}[t]
\centering
\includegraphics[width=\linewidth]{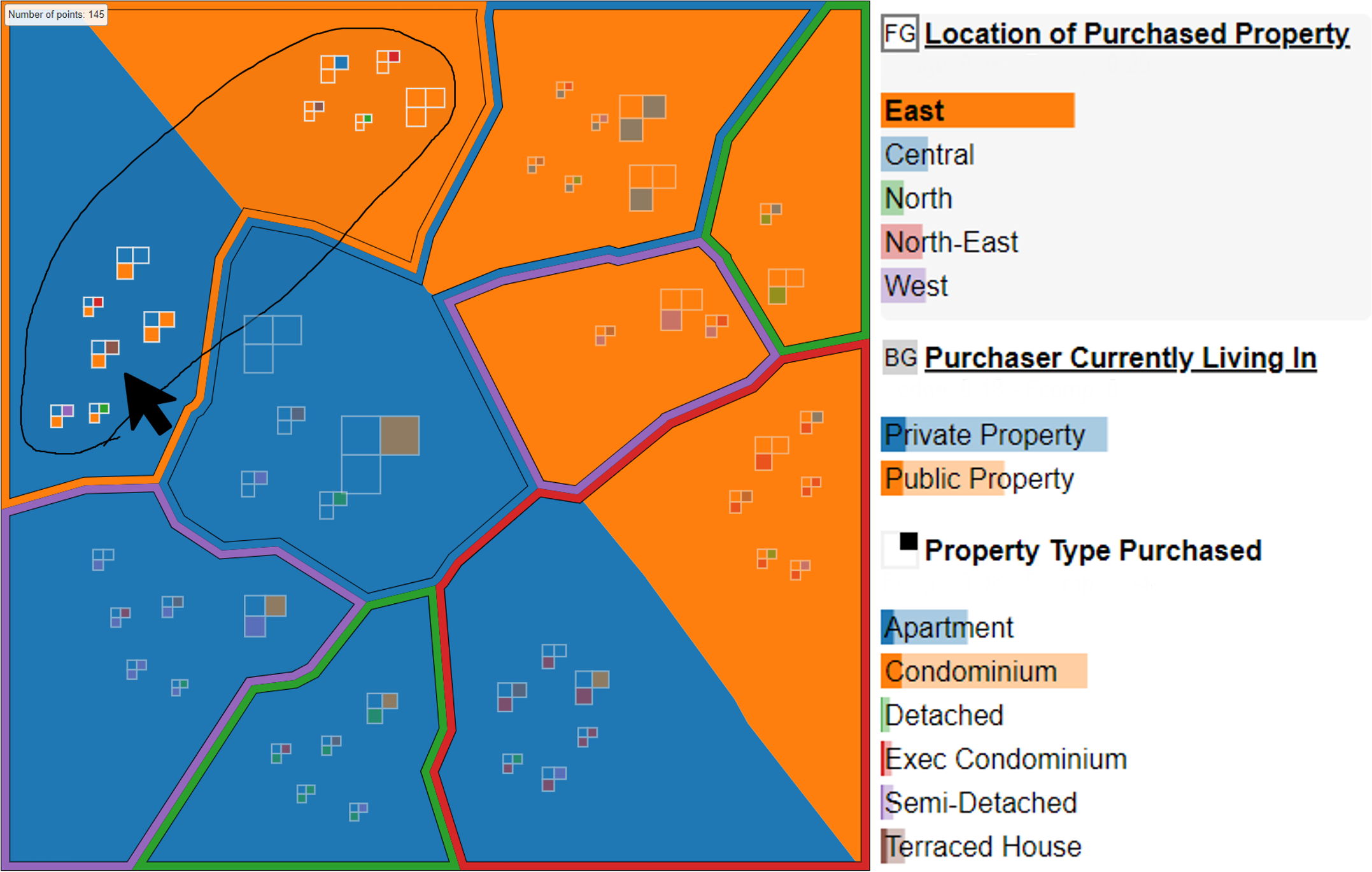}
\caption{
Through user selection, the borders of a second attribute can be added to the foreground of the plot, e.g., \textrm{Purchaser Currently Living In} is shown in the background as the primary attribute, and \textrm{Location of Purchased Property} is shown in the foreground.
}
\label{fig:selection}
\end{figure}

\subhead{Attribute Selection:}
Users can change the attribute visualized through layout enrichment.
We also show the outline for categories of a second selected attribute (see~\autoref{fig:selection}).
We add the borders of categories to the foreground if another attribute is already selected and visualized in the background.
This visual cue does allow for the observation of one main attribute and a second attribute, similar to the outline of MosaicSets~\cite{Rottmann2023}.
This introduces less clutter and thus requires less effort to perceive.
We initially used textures with different colors to represent different categories.
However, using textures of different colors to fill each cell in the Voronoi portioning introduced excessive clutter, and the interpretation of common regions was difficult.

\subhead{Subset Selection:}
We allow for the selection and highlighting of groups of subsets.
Once the user has selected data items, we show the common categories of the selection using Lasso selection and highlight all data items outside of the selection with the same combination of categories in the side panel on the left, similar to the proximity visualization for continuous data proposed by Aupetit and Catz \cite{Aupetit2007}.
This interaction enables cluster analysis since all common categories among the selected items are highlighted (see side panel in \autoref{fig:selection}).
Thus, visual groupings can be compared with respect to the categories and attributes contributing to cluster cohesion.
Additionally, all subsets matching the common categories of the selection are also highlighted (see plot in \autoref{fig:selection}).
Together, this allows analysts to observe and judge group cohesion along with the contributing attributes.

\subhead{Attribute and Category Ordering:}
A user can select attributes of the dataset listed on the side panel to change the attribute encoded in the foreground and background of the plot.
By default, attributes are sorted by their edge-based fracturedness in ascending order, and categories are ordered by their individual contributions to component-based fracturedness in ascending order, allowing for a focus on attributes forming clear splits in the projection space.
When selecting subsets (see previous paragraph), the lists of common attributes and distinct attributes are also ordered similarly.

\section{Interpreting the Categorical Data Map}

In the following, we perform a case study on cluster and attribute analysis, using the Property Sales dataset~\cite{Koh2011} (see~\autoref{fig:basic-example}) to show how to interpret emerging patterns for cluster, outlier, and similarity analysis.
We chose this dataset because of its relative simplicity.
However, it lacks the complexity of large categorical datasets, which we will address in an expert study (see \autoref{sec:evaluation}).

\subhead{Cluster Analysis:}
\label{sec:evaluation:cluster-analysis}
There exist a total of $\Pi_{n \in \{1,\dots,|\mathcal{A}|\}} \, \left|\mathcal{C}_n\right|$ possible data items, given that all combinations of attributes are allowed, resulting in an exponential growth in the number of possible and unique data items.
Hence, we can assume that there are dependencies and relationships among the categories contained in a dataset impacting their distribution.
This means that groups of subsets that share a set of attributes should form perceivable structures (i.e., clusters) when projected using DR methods.
Thus, our approach benefits from and leverages the sparsity of categorical data.
For the Property Sales dataset, there are ten clusters (see~\autoref{fig:basic-example} (1)).
There is a symmetric split along the center of the projection.
Given the size of this dataset, we can observe that the two attributes \textsf{Purchaser Currently Living In} and \textsf{Location of Purchased Property} dominate the appearance of the projection.
The glyph sizes indicate that the categories \textsf{Private Propriety} and \textsf{Central} often occur together while \{\textsf{Private Propriety}, \textsf{Central}, \textsf{Condominium}\} is the largest unique subset.
Thus, we can see that most private property is purchased in the central areas, and in this general group, the large majority are condominiums.

\begin{table*}
\setlength{\tabcolsep}{2.2mm}
{
\centering
\begin{tabular}[c]{lcccccccccccccccccc}
\toprule
\multirow{2}{*}{\textbf{Dataset}} & \multicolumn{3}{c}{\textbf{TW} ($\uparrow$)} & \multicolumn{3}{c}{\textbf{CT} ($\uparrow$)} & \multicolumn{3}{c}{\textbf{SC} ($\uparrow$)} & \multicolumn{3}{c}{\textbf{NS} ($\downarrow$)} & \multicolumn{3}{c}{\textbf{Avg. NH} ($\uparrow$)} & \multicolumn{3}{c}{\textbf{Med. NH} ($\uparrow$)} \\
\cmidrule(lr){2-4} \cmidrule(lr){5-7} \cmidrule(lr){8-10} \cmidrule(lr){11-13} \cmidrule(lr){14-16} \cmidrule(lr){17-19}
& \rotatebox{90}{\textbf{MDS+O}} & \rotatebox{90}{\textbf{MDS+J}} & \rotatebox{90}{\textbf{MCA}} & \rotatebox{90}{\textbf{MDS+O}} & \rotatebox{90}{\textbf{MDS+J}} & \rotatebox{90}{\textbf{MCA}} & \rotatebox{90}{\textbf{MDS+O}} & \rotatebox{90}{\textbf{MDS+J}} & \rotatebox{90}{\textbf{MCA}} & \rotatebox{90}{\textbf{MDS+O}} & \rotatebox{90}{\textbf{MDS+J}} & \rotatebox{90}{\textbf{MCA}} & \rotatebox{90}{\textbf{MDS+O}} & \rotatebox{90}{\textbf{MDS+J}} & \rotatebox{90}{\textbf{MCA}} & \rotatebox{90}{\textbf{MDS+O}} & \rotatebox{90}{\textbf{MDS+J}} & \rotatebox{90}{\textbf{MCA}} \\
\midrule
\textbf{Audiology} \cite{Bareiss1988} & .58 & \textbf{.89} & .83 & .64 & \textbf{.90} & .89 & .29 & \textbf{.77} & .69 & .17 & \textbf{.09} & .81 & .89 & \textbf{.92} & \textbf{.92} & .98 & .98 & .98 \\
\textbf{Mushroom} \cite{Lincoff1981} & .96 & \textbf{.97} & .91 & .92 & .93 & \textbf{.97} & .78 & .77 & \textbf{.79} & \textbf{.08} & .09 & .64 & .89 & \textbf{.90} & .84 & .90 & \textbf{.92} & .87 \\
\textbf{Titanic} \cite{Titanic95} & \textbf{.86} & \textbf{.86} & .76 & \textbf{.84} & \textbf{.84} & .81 & \textbf{.76} & .75 & .59 & \textbf{.07} & .07 & .28 & \textbf{.68} & \textbf{.68} & .63 & .74 & \textbf{.75} & .60 \\ 
\textbf{Cyber-Security} \cite{Yano15} & .84 & \textbf{.87} & .79 & .83 & \textbf{.86} & .81 & \textbf{.87} & .82 & .68 & \textbf{.04} & .06 & .30 & \textbf{.56} & .55 & .54 &\textbf{.66} & .62 & .58 \\
\textbf{Property Sales} \cite{Koh2011} & \textbf{.91} & .89 & .73 & \textbf{.86} & .85 & .81 & \textbf{.70} & .66 & .46 & \textbf{.09} & .10 & .24 & \textbf{.65} & \textbf{.65} & .51 & \textbf{.76} & \textbf{.76} & .39 \\
\textbf{HCI Study (1)} \cite{Rogers16} & \textbf{.84} & .77 & .71 & \textbf{.81} & .79 & .73 & \textbf{.72} & .69 & .61 & \textbf{.07} & \textbf{.07} & .18 & \textbf{.59} & .58 & .58 & \textbf{.53} & \textbf{.53} & .52 \\
\textbf{HCI Study (2)} \cite{Rogers16} & 1.0 & 1.0 & 1.0 & 1.0 & 1.0 & 1.0 & .86 & .85 & \textbf{.89} & \textbf{.03} & \textbf{.03} & .11 & .56 & .56 & .56 & .57 & .57 & .57 \\ 
\bottomrule
\end{tabular}
}
\caption{We compare projections of MDS using the Overlap coefficient (\textit{MDS+O}) and the Jaccard distance (\textit{MDS+J}) to \textit{MCA} by applying them to seven real-world datasets. The \textit{MDS} outperforms \textit{MCA} for most datasets and quality metrics. In the case of the \textit{Audiology} dataset with high category overlap, usually present in datasets with many attributes, we found that \textit{MDS} combined with the Jaccard distance outperforms both alternatives.}
\label{tab:evaluation-metrics}
\end{table*}

\subhead{Attribute Analysis:}
\label{sec:evaluation:attribute-analysis}
By encoding the attribute values in the background, we enable users to analyze the distribution of subsets in the projection with respect to one or two attributes.
For the Property Sales dataset, we found that the attribute \textsf{Purchaser Currently Living In} creates a clear and straight division between subsets (see~\autoref{fig:basic-example} (2)).
We can also see a second level of grouping by the \textsf{Location of Purchased Property} attribute forming a close to orthogonal split in the projection, which can be spotted with our visualizations (see~\autoref{fig:basic-example} (3) and \autoref{fig:selection}).
Thus, \textsf{Purchaser Currently Living In} and \textsf{Location of Purchased Property} are the primary attributes.
This finding is substantiated when checking the side panel entry of the attribute \textsf{Property Type Purchased}, which has three categories with low frequency. 
The appearance of the partitioning depends a lot on the selected attributes.
When observing the layout enrichment, attributes present themselves on a spectrum from a few clearly separated groups to intermingled and highly fractured appearances.
\textsf{Property Type Purchased} does not contribute to elements' clustering (or cluster cohesion) since most groups contain subsets of the majority of its categories.
Thus, the areas of the categories are disjointed, which reflects the fact that most property types are sold as both private and public property, as well as most of the geographic locations.

\begin{figure}
\centering
\includegraphics[width=0.49\linewidth]{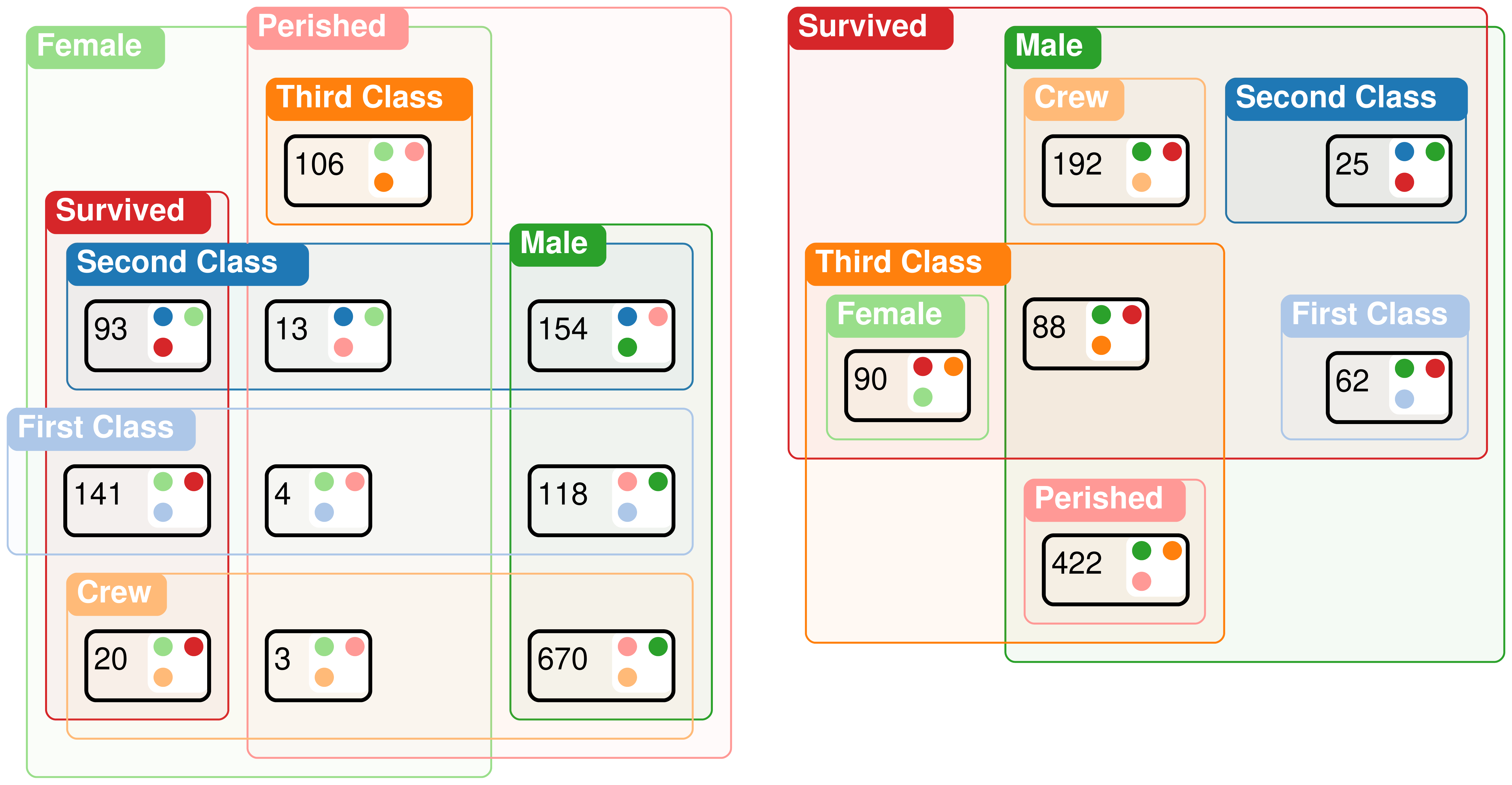}\hspace{0.5cm}
\includegraphics[width=0.42\linewidth]{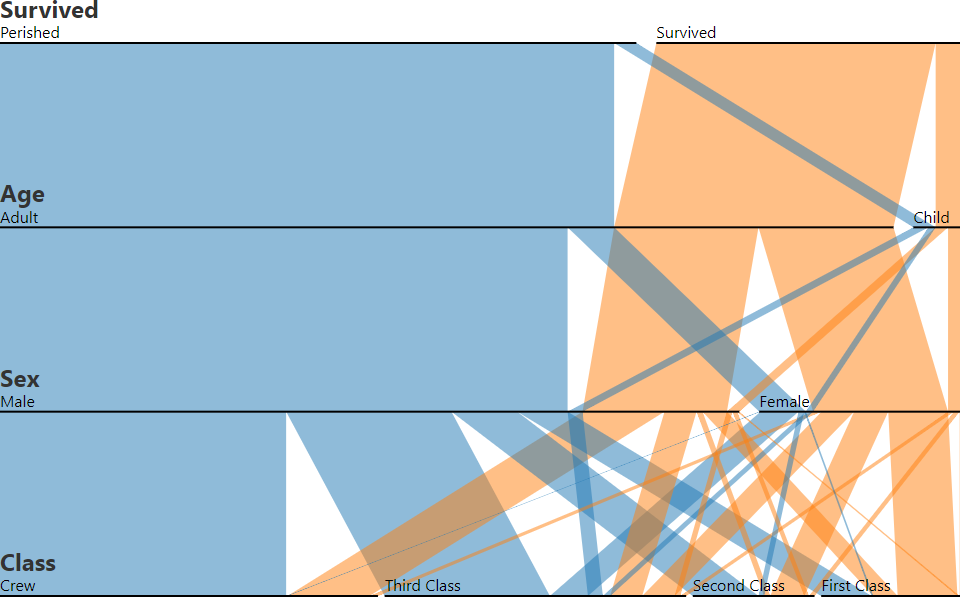}\\
\caption{
Two visualizations of the Titanic dataset \cite{Titanic95}.
A \emph{split} Euler diagram without the \textsf{Age} attribute (left) and an overlap reduced Parallel Sets visualization (right) with \emph{very thin ribbons}. 
Both have drawbacks with a small dataset and do not scale with an increasing number of attributes. 
}
\label{fig:set-vis-examples-eval}
\end{figure}

\begin{figure*}
\includegraphics[width=\textwidth]{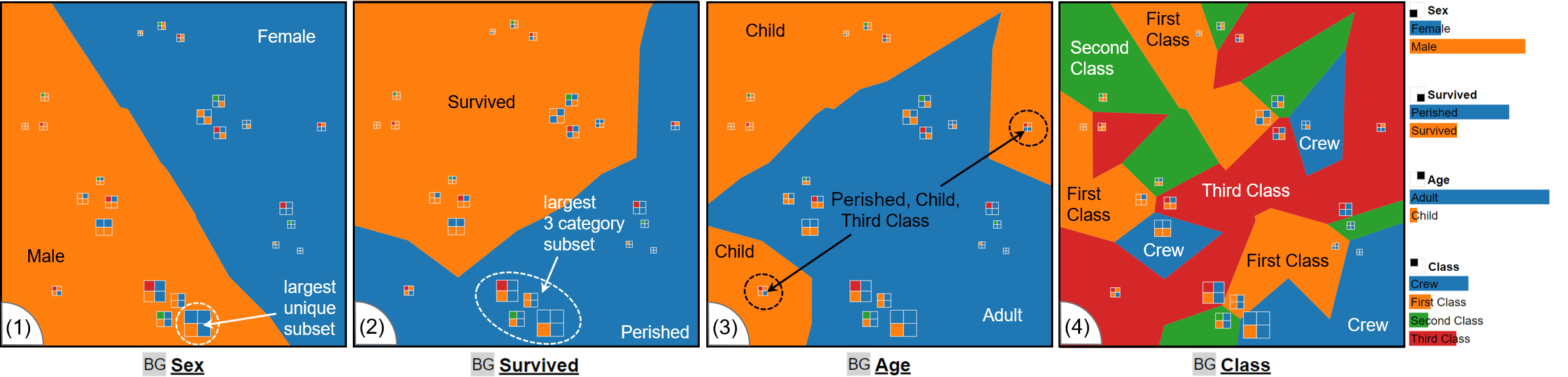}
\caption{
\emph{Categorical Data Map} visualizations of the \emph{Titanic dataset}~\cite{Titanic95} using \emph{MDS} and \emph{Overlap coefficient}~\cite{Szymkiewicz1934}.
(1) The visualization shows six clusters and two outliers.
The largest cluster is the subset of \textsf{Adult, Male, Perished} (at the bottom).
The background encoding shows that the \textsf{Survived} and \textsf{Sex} attributes are relevant for this dataset, clearly separating the data items.
For \textsf{Sex}, the separation is left and right.
(2) For \textsf{Survived}, the separation is bottom-right/top-left.
(3) The \textsf{Age} dimension also yields a separation, while (4) \textsf{Class} shows no clear structure.
}
\label{fig:titanic}
\vspace*{-1em}
\end{figure*}

\section{Evaluation}
\label{sec:evaluation}

We qualitatively compare our \emph{Categorical Data Map} to existing visualizations for categorical data and quantitatively compare our approach to MCA used by Broeksema et al. \cite{Broeksema2013}.
Additionally, we performed an expert study on two representative datasets with five data scientists.

\subsection{Comparison to Euler Diagrams and Parallel Sets}

For categorical data, each data point has exactly one category for each attribute, while in Euler diagrams, the number of sets an element is included in is not restricted, i.e., it could be in less.
Thus, to truthfully represent categorical data in Euler diagrams, there need to be $\Sigma_{a_i \in \mathcal{A}} \,|\mathcal{C}_i|$ sets, i.e., one set for each category of all attributes.
Euler diagrams may require the selection of specific subsets of attributes and, therefore, are less suitable for exploratory data analysis.
For highly intersecting sets, automatic layout methods might not create a single diagram~\cite{Paetzold2023}. 
We show an example of an automatically generated split Euler diagram for the Titanic dataset in~\autoref{fig:set-vis-examples-eval} (left).
The attribute \textsf{Age} was removed to reduce the diagram's complexity.
The Titanic dataset requires ten sets.
However, even with eight sets, the visualization is disjointed.
Parallel Sets are alternative categorical sets visualization, combining principles from stacked bars and parallel coordinate plots \cite{Bendix2005,Kosara2006}.
\autoref{fig:set-vis-examples-eval} (right) shows the Titanic dataset in a Parallel Sets visualization, where the readability is improved through overlap reduction.
Small subsets are represented as very thin ribbons on the lowest level, which can be hard to perceive.
Visualizing the Mushroom dataset with classical Parallel Sets is not visually feasible since it will have 22 ribbon layers and 8123 subsets on the lowest level (see supplementary material).
Alsakran et al. ~\cite{Alsakran2014} addressed this issue by only visualizing 2-dimensional subsets in a modified Parallel Sets visualization.
However, the relation between 2-dimensional subsets is lost.
Thus, we argue that Euler diagrams and Parallel Sets, as examples of established visualizations for categorical data, do not scale with an increasing number of attributes.

\subsection{Quantitative Evaluation of Projection Quality}

We use five quality metrics commonly used in related work for DR to evaluate and compare the quality of our categorical data projections \cite{Espadoto2021}.
The result of comparing MDS with Overlap coefficient (\textit{MDS+O}) and Jaccard distance (\textit{MDS+J}) to \textit{MCA} are shown in \autoref{tab:evaluation-metrics}.
We briefly describe each metric below and use them to compare our MDS-based method to MCA using seven real-world categorical datasets.

\subhead{Trustworthiness (TW)} \cite{Venna2006} quantifies the proportion of points that remain close in the lower-dimensional representation to assess how accurately local patterns in the projection represent the data patterns.
This is linked to the occurrence of "false neighbors" in the protection.
\subhead{Continuity (CT)} \cite{Venna2006} measures the ratio of points in the projection that remain close in the original space.
This is related to the "missing neighbors" of a projected point.
\subhead{Normalized Stress (NS)} \cite{Joia2011} quantifies how well the distances between pairs of points are preserved when mapping from the original space to the projected space. This measure should be as low as possible.

\subhead{Shepard Diagram Correlation (SC)} \cite{Espadoto2020} measures the rank correlation of all distances of the original and the projected space, assessing the quality of distance preservation globally using Spearman's $\rho$ \cite{Spearman1904}.

\subhead{Neighborhood Hit (NH)} \cite{Paulovich2008} measures the proportion of a point's neighbors in the projection space that share the same label as the point itself, averaged across all points in its neighborhood. This metric is related to the separation of labeled data in the projection.
In our case, we evaluate every attribute as a set of labels.
Thus, we calculate the mean and median values of \textbf{NH} across all attributes of a dataset.

\smallskip

\textbf{TW}, \textbf{CT}, \textbf{NH} require a parameter $k$ defining a neighborhood size. We set $k = 7$, a commonly used value \cite{Espadoto2020}.
We found that our approach generally outperforms \textit{MCA} quantitatively.
Additional measurements of MDS with other distance measures, detailed descriptions of the quality metrics, and datasets are provided in the supplementary material.

\subsection{Qualitative Expert User Study}

To evaluate the \emph{Categorical Data Map} we performed a paired analytics study \cite{Kaastra2014}.
We conducted an expert study with five data scientists, \textbf{E1--E5}, with varying backgrounds.
All participants were Ph.D. candidates and students.
All were male, and the age range was 25 to 30 years.
All experts had experience in the area of information visualization and visual analytics.
During the study, we asked the experts to verbalize their thought process to capture it.
The following studies are set up using MDS projections of the Mushroom and Titanic dataset using the Overlap coefficient (\textit{MDS+0}). \autoref{tab:evaluation-metrics} shows that these projections are higher quality than MCA-based ones regarding most quality metrics.

All trials followed a predefined structure and took between 43 and 57 minutes.
The study was conducted in German.
The study started with an introduction to the \emph{Categorical Data Map} using the Property Sales dataset by Hassan et al.~\cite{Hassan2014} shown in~\autoref{fig:basic-example} and included a description of the square area glyph, layout enrichment, and interactions to introduce the expert to the prototype.
After the introduction, the experts had the opportunity to ask questions regarding our approach.

\subhead{Titanic Dataset:}
The experts had to analyze the Titanic dataset~\cite{Titanic95} using the \emph{Categorical Data Map} shown in~\autoref{fig:titanic}.
\textbf{E1--E5} were able to locate the largest subset \{\textsf{Male}, \textsf{Perished}, \textsf{Adult}, \textsf{Crew}\} by looking at the visualization without any additional interaction (\autoref{fig:titanic} (1)).
\textbf{E1--E5} used Lasso selection to find and validate that the largest subset regarding three attributes is \{\textsf{Male}, \textsf{Perished}, \textsf{Adult}\} (\autoref{fig:titanic} (2)).
Additionally, \textbf{E1--E5} were able to find six clusters and two outliers.
\textbf{E1}, \textbf{E3}, and \textbf{E4} found that the outliers represent the subsets defined by the categories \{\textsf{Perished}, \textsf{Child}, \textsf{Crew}\} (\autoref{fig:titanic} (3)).
\textbf{E1}, \textbf{E3}, and \textbf{E5} commented on the high number of perished males and the large number of casualties among the \{\textsf{Male}, \textsf{Crew}\}.
\textbf{E1--E5} used the layout enrichment to navigate and reason about the location of subsets, including the \textsf{Class} attribute (\autoref{fig:titanic} (4)).
\textbf{E2} commented on the close to orthogonal split in the projection between \textsf{Sex} and \textsf{Survived} shown in \autoref{fig:titanic} (1) and (2).

\begin{figure*}[t]
\centering
\includegraphics[width=\linewidth]{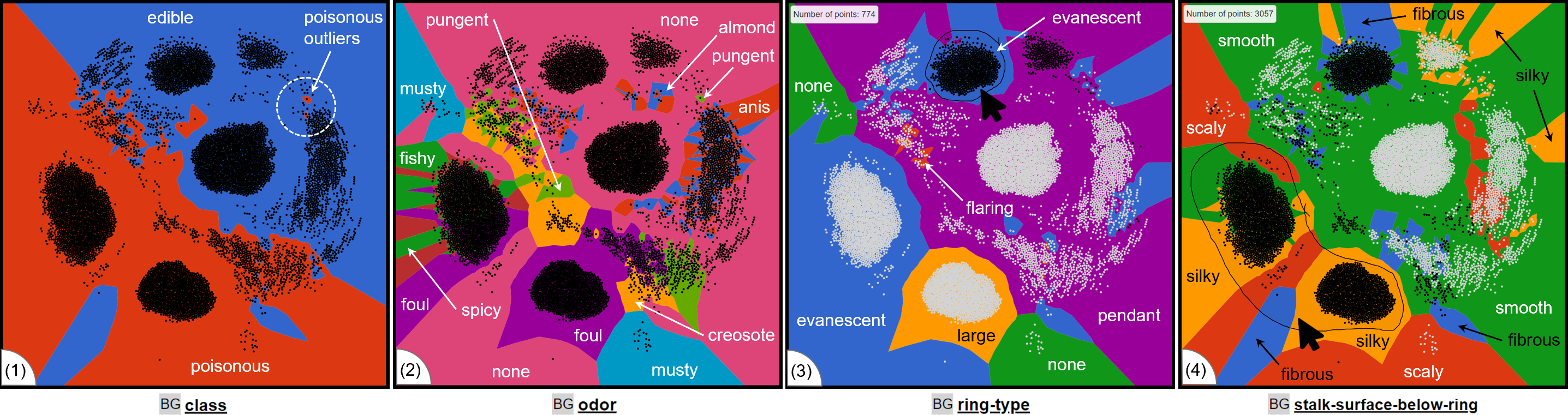}
\caption{
\emph{Categorical Data Map} visualizations of the \emph{Mushroom dataset} \cite{Lincoff1981} using the \emph{MDS} and \emph{Overlap coefficient}.
(1) Two \textsf{poisonous} mushrooms very similar to \textsf{edible} mushrooms. 
(2) Comparing \textsf{class} and \textsf{odor} reveals that the poisonous outlier has a \textsf{pungent} odor.
Continued analysis reveals that mushrooms with an unpleasant smell are \textsf{poisonous}.
(3) After the selection of a cluster, the \textsf{ring-type} is identified as a defining characteristic for the cohesion of visible clusters and is used as a property for the classification of mushrooms.
(4) Selecting two \textsf{poisonous} clusters, reveals that the vast amount of \textsf{poisonous} mushrooms are \textsf{silky} at the \textsf{stalk-surface-below-ring}, while there exist very few \textsf{silky} mushrooms that are \textsf{edible}.
}
\label{fig:mushroom}
\end{figure*}

\subhead{Mushroom Dataset:}
\textbf{E1--E5} had the opportunity to perform an open exploration task and were only given the information that the dataset is about mushrooms and that the \textsf{class} attribute indicates their \textsf{poisonousness}.
The glyph was replaced with a simple black dot to reduce the visual complexity.
\textbf{E1--E5} perceived five clusters right at the outset. 
\textbf{E2} and \textbf{E3} used the Lasso selection together with layout enrichment to determine differentiating categories for cluster separation, e.g., \textsf{evanescent}, \textsf{large}, and \textsf{pendent} for the \textsf{ring-type} attribute (\autoref{fig:mushroom} (3)).
\textbf{E1--E5} found the \textsf{poisonous} outliers nested in the group representing \textsf{edible} mushrooms (\autoref{fig:mushroom} (1)) being poisonous mushrooms very similar to edible ones.
\textbf{E1--E5} found the general rule that mushrooms with an \textsf{fishy} 
\textsf{foul}, \textsf{musty}, \textsf{spicy} or other unpleasant smells indicate a \textsf{poisonous} mushroom (\autoref{fig:mushroom} (2)).
During the open exploration task, \textbf{E1}, \textbf{E2}, and \textbf{E3} found the rule without additional information.
\textbf{E4} and \textbf{E5} needed help to find the \textsf{class} and \textsf{odor} combination.
However, \textbf{E4} and \textbf{E5} could deduce the rule by only interpreting the plot.
\textbf{E3}, quickest in exploring the dataset, found that \textsf{stalk-surface-below-ring} is \textsf{silky} for the majority of \textsf{poisonous} mushrooms and the \textsf{stalk-surface-below-ring} is mostly \textsf{smooth} for \textsf{edible} ones (\autoref{fig:mushroom} (4)).

\subhead{General Comments:}
Before concluding the study, the participants were asked to comment on their preferences for the available glyph designs.
\textbf{E1}, \textbf{E2}, and \textbf{E4} preferred a circular glyph design (\autoref{fig:glyphs} (c) and (d)) over a square design.
\textbf{E3} and \textbf{E5} preferred square glyph designs (\autoref{fig:glyphs} (a) and (b)).
\textbf{E1} and \textbf{E3} found that the area-based glyphs are inferior to the alternative designs for reading off precise subset sizes.
\textbf{E1} mentioned as a drawback that the glyphs are not rotation invariant.
\textbf{E1} commented that the layout enrichment is very useful for navigation and orientation and helps to perceive the impact on category groups.
However, \textbf{E1} also noted that the layout enrichment does not reflect the ratio of data items with a given category.
\textbf{E3} mentioned a general preference for the map metaphor by being helpful for orientation among different subsets.
\textbf{E2} mentioned potential scalability issues with the glyph for large datasets, e.g., for a high number of attributes, and proposed semantic zoom as a potential option.
\textbf{E1--E5} commented that ordering attributes according to their fracturedness was understandable and useful.
During the general questions at the end, \textbf{E2--E4} freely explored plots created with other distance measures and DR methods.
\textbf{E3} commented that the result of MCA-based plots was hard to interpret, noticing the disjointed layout enrichment and thus having larger fracturedness.
\textbf{E1} mentioned issues with the encoding of categories, such as the category \textsf{North} not being located north of the plot or the category \textsf{brown} not having the color brown, and suggested being able to select the color of a category manually.

\section{Discussion and Future Work}
\label{sec:discussion}

In this section, we discuss the lessons-learned, reflect on the design decisions, and discuss computational complexity and future work.

\subhead{Visualizing Attributes and Categories:}
We initially used circular glyphs as shown in \autoref{fig:glyphs} (c) and (d), which had the benefit of using the available space effectively since overlap minimization relative to the radius is straightforward to implement.
The subset size encoding by the arc around the circle enables finer-grained distinction of sizes since it offers more space.
However, during the design phase, users misinterpreted the circle segments as pie charts, a common method for displaying categorical data.
Thus, we decided to circumvent this common misconception by using square-based representation for the categorical subsets.
However, three out of five experts preferred a circular glyph design.

There are visual limitations to the number of dimensions and categories that our approach is able to support.
The number of visually distinguishable categories is limited by the number of square segments that fit into the glyph, which is limited by the screen space.
The number of attributes is limited by the number of colors, which have to be distinguishable and memorizable.
Thus, we suggest following Miller's Law~\cite{Miller1956} for the number of dimensions and attributes, which proposes a maximum of seven plus or minus two.
Alternatively, we suggest interactions such as semantic zoom, e.g., removing attributes for which all subsets have the same category after zooming in on a specific area.
\subhead{Encoding of Subset Sizes:}
We evaluated four different visual encodings for the size of a categorical data subset (see \autoref{fig:glyphs}).
The area-based glyph makes it easier to perceive subset sizes at a glance, and thus, a user can spot the distribution of the dataset directly.
Still, it suffers from overlap, especially for tight clusters.
Thus, there is a benefit to applying methods to reduce overlap.
We are able to mitigate the overlap problem with the force-directed overlap reduction largely.
Simplifying the representation of a dot requires less space, but the assessment of subset sizes requires interaction.
It is possible to remove the subset size information altogether.
However, this may limit analysis tasks where the subset sizes it not important, e.g., the Mushroom dataset.
All glyph designs benefit from a mouse-over mechanism that moves the currently selected glyph to the top so that all attributes can be observed.
\subhead{Encoding of Attributes Into the Background:}
\autoref{fig:titanic} shows that encoding an attribute into the visualization gives insight into the topology of the projection.
We could also show the benefit of encoding multiple attributes into the background to allow for a more complex representation of the topology.
We found that the number of categories of an attribute weakly influences the fracturedness of an attribute.
However, the main factor is the number of subsets containing the attribute, i.e., an attribute with two categories and an occurrence roughly equal among all subsets will yield a low fracturedness for that attribute.
With increased imbalance between the categories, the fracturedness may increase if other more balanced attributes are present. 
We discussed the use of \emph{weighted Voronoi diagrams}~\cite{Ash1986} to better reflect the subset size in the background encoding.
The use of a weighted Voronoi diagram will conflict with local cluster patterns; more specifically, for imbalanced datasets, the area of one Voronoi cell extends below the point of its neighbors, requiring restrictions on the range weights.
This behavior makes the layout enrichment hard to interpret since points are placed inside or close to an area representing a category they do not belong to.
For datasets with only unique entries, the weight Voronoi diagram will be identical to the regular Voronoi diagram.
To organize subsets, we also considered \emph{Voronoi Treemaps}~\cite{Balzer2005}.
However, Voronoi Treemaps require a hierarchical structure, just like regular Tree Maps~\cite{Shneiderman1992} and, thus, cannot be applied to categorical data without additional information to derive a hierarchy of attributes.

\subhead{Computational Complexity:}
The number of data records $n$ poses potential limitations.
The time complexity of projecting data is determined by the DR methods.
However, since categorical data sets are sparse, as discussed in~\autoref{sec:evaluation:cluster-analysis}, the number of projected subsets is significantly lower than that of data records.
The Voronoi diagram calculation and the corresponding Delaunay triangulation are both in $O(n~log(n))$ \cite{Aurenhammer1991}.
The time complexity of calculating the fracturedness measures depends on the number of vertices and edges of the Delaunay triangulation, which will have $n$ vertices and $3n-3-h$ edges, where $h$ is the number of vertices on the convex hull.
The time complexity of calculating \emph{edge-based fracturedness} is based on enumerating all edges of the Delaunay triangulation and has a time complexity of $O(|E|)$.
The time complexity of calculating \emph{component-based fracturedness} is dependent on the algorithm for determining the number of components.
We use a depth-first search-based approach with time complexity of $O(|V| + |E|)$.
Thus, the dimensionality reduction method employed poses the highest contribution to the time complexity, $O(n^3)$ for MDS.
%
%and for t-SNE $O(n^2)$, which will be the overall time complexity.
%

\subhead{Future Work:}
We found that Voronoi cells can overrepresent the amount of data associated with a specific category.
Thus, there is a need for a new layout enrichment method following these constraints: (1) The global area associated with one category should be relative to the occurrence in the dataset (data-ink ratio), (2) the extent of individual category areas should remain close to their projected data point positions, (3) where meaningful (e.g., among clusters), the layout enrichment should visually enclose the data points with the same category if the data-ink ratio allows.
The expert study showed that the color assignment for foreground and background colors could be improved.
We suggest assigning attributes to a few sets of colors based on an exploration phase.
Later in the analysis, we require one color set for the attribute used in the background, one for the foreground using a distinctive palette, and one for the attribute under focus by the user.
All the other attributes would be assigned a neutral color (e.g., grey).
In this paper, we studied the use of MDS for categorical data analysis.
However, following the approach of encoding categorical data into distances, other DR methods could be used (e.g., t-SNE \cite{Maaten2008} or UMAP \cite{Mcinnes2020}).
These can be evaluated and compared quantitatively, following the evaluation presented in this paper.
We think that the concept of \emph{fracturedness} can be transferred to high-dimensional space when analyzing categorical data.
Such a measure can be used to compare the low- and high-dimensional representations and provide a quality measure for projections of categorical data.

\section{Conclusion}
\label{sec:conculsion}

We presented a novel projection-based visualization method to address the need for similarity-based analysis techniques for categorical data.
We leverage distance relations based on set intersections to create enhanced and interactive glyph-based scatterplot-like visualizations called the \emph{Categorical Data Map}.
We visualized attributes and categories by calculating a Voronoi partitioning and coloring the cells according to the category of the associated attribute. 
Our method allows for exploring the categorical data space through segmentation, enabling the orientation along an automatic or user-selected attribute.
For automatic selection, we rank-order attributes along a visual property we defined as \emph{fracturedness} measures. 
We quantitively evaluated different distance measures for the projection of categorical data with MDS, suggesting that the Overlap coefficient and Jaccard distance yield results outperforming MCA.
Through a case study, we showed that our \emph{Categorical Data Map} can support the identification of similar subsets and clusters, as well as the detection of attributes with a strong influence on the topology of the embedding.
In an expert study, we were able to confirm that our approach facilitates the analysis of categorical data, especially for large datasets, by grouping similar subsets while, through layout enrichment, visualizing the distribution of categories of an attribute. 
We published a demonstrator and our results online so that users can interactively experiment with our approach and build upon our results.
We conclude that the \emph{Categorical Data Map} effectively analyzes large categorical datasets, especially in exploratory scenarios.

%% if specified like this the section will be omitted in review mode
\acknowledgments{%
%\section{Acknowledgments}

We thank the anonymous reviewers for their valuable feedback.
This work was funded by the Deutsche Forschungsgemeinschaft (DFG, German Research Foundation) -- Project-ID 251654672 -- TRR 161 (Projects A01 and A03) and the Federal Ministry for Economic Affairs and Climate Action (BMWK, grant No. 03EI1048D).

}

\bibliographystyle{abbrv-doi-narrow}

\bibliography{00-references}

\end{document}